\documentclass{elsart}
\usepackage{graphicx,amsmath,amsfonts, ecltree, epic, amsmath, amssymb, amsfonts}

\begin{document}
\begin{frontmatter}

\title{Kinetic formulation of irreversible evolution of the two-nucleus spin systems
}
\author{S.Eh.Shirmovsky}
\ead{shirmov@ifit.phys.dvgu.ru} \address{ Laboratory of Theoretical Nuclear Physics, Far Eastern Federal University,
8 Sukhanov St., Vladivostok, 690950, Russia\\
Physics, mathematics and informatics cathedra, Vladivostok State Medical University, 
2 Ostryakov St., Vladivostok, Russia}
\date{\today}

\begin{abstract}
The time irreversible evolution of the two-nucleus spin systems interacting with a magnetic field is analyzed in the framework of the subdynamics theory. The spin systems are determined by the H(1) and C(13) nuclei.
The investigation is based on the complex spectral representation of the Liouvillian which gives a rigorous description of irreversibility.  The evolution of the density matrix is analyzed for the short and long time region. Quantum Zeno effect is discussed. 
\end{abstract}
\end{frontmatter}

\section{Introduction}
The irreversible nature of the world impels us to develop theory by adequate means, in accordance with this fact.
However,  the exact description of irreversibility on the basis of
fundamental classical and quantum theories  is impossible,  since they are reversible in the time.  
As a rule, the problem is solved on the basis of different physical approximations such as the Pauli master equation~\cite{blum},
the Lindblad axiomatic formalism~\cite{Isar}. In these cases coarse-graining approximations, the distinction between an 
open system and its environment are used. The environment is assumed to be in thermodynamic equlibrium. Also,
the problem is solved on the basis of the Schroedinger equation with a dissipative term~\cite{kad}, on the basis of the mixed quantun-classical approaches~\cite{lachno}. 
On the whole, the approaches mentioned above, having a nature of the approximations, contain a subjective, anthropomorphic element and cannot be completely satisfactory.
Thus, the problem of describing the irreversible world on the basis of
the reversible equations of classical and quantum physics i.e., the determination of the objective laws of nature arises. \\
It is obvious that fundamental solution to this problem can be achieved when it is based on the equations which originally have irreversible nature. 
In this case, one should speak about the  formulation of the dynamics, which makes it possible to include  irreversibility in a natural way.\\
For many years, the Brussels-Austin group, that was headed by I. Prigogine, has been developing an alternative approach with the aim to reveal the role of the irreversible processes, as well as their fundamental role in nature.
A conclusion has been made that the solution to this task may be found outside the Hilbert space. In this approach, the eigenvalues problem is reduced to the complex eigenvalues problem~\cite{liuv1},~\cite{ppt}
\begin{align}\label{ht1}
H
\mid\varphi_{\alpha}>=Z_{\alpha}\mid\varphi_{\alpha}>,~<\widetilde{\varphi}_{\alpha}\mid
H = <\widetilde{\varphi}_{\alpha}\mid Z_{\alpha},
\end{align}
where the right eigenstate $\mid\varphi_{\alpha}>$ and the left eigenstate $<\widetilde{\varphi}_{\alpha}\mid$ must be distinguished and
the  eigenvalue $Z_{\alpha}$ is a complex number 
\begin{equation}\label{gamma}
Z_{\alpha}={\bar{E}_{\alpha}} - i\gamma_{\alpha}.
\end{equation}
Here $\bar{E}_{\alpha}$ is a renormalized energy and $\gamma_{\alpha}$ is a real
positive value. The index $\alpha$ corresponds to the certain quantum state.
In this case, the eigenstates $\mid\varphi_{\alpha}>$,
$<{\tilde{\varphi}_{\alpha}}\mid$ have a broken time symmetry. The eigenstate $\mid\varphi_{\alpha}>$ is associated with the unstable state
which vanishes for $t\rightarrow +\infty$, while $\mid{\tilde{\varphi}_{\alpha}}>$ corresponds to the state vanishing for
$t\rightarrow - \infty$
\begin{equation}\label{hq}
\mid\varphi_{\alpha}(t)>=\exp(-i{\bar{E}_{\alpha}}t-\gamma_{\alpha}
t)\mid\varphi_{\alpha}(0)>,
\end{equation}
\begin{equation}\label{hq}
\mid\tilde{\varphi}_{\alpha}(t)>=\exp(-i{\bar{E}_{\alpha}}t+\gamma_{\alpha}
t)\mid\tilde{\varphi}_{\alpha}(0)>.
\end{equation}
The eigenstates $\mid\varphi_{\alpha}>$, $<\widetilde{\varphi}_{\alpha}\mid $ are named the Gamow
vectors and are in detail  discussed in the literature
~\cite{gam1}~-~\cite{gam4}.\\
Further development of the Brussels-Austin group approach was obtained on the basis of the Liouville space extension 
of quantum mechanics~\cite{liuv1},~\cite{liuv2}, in the framework of the complex spectral representation of Liouville-von Neumann operator~$L$ (Liouvillian) and subdynamics  theory.   In this case, Liouvillian has complex eigenvalues that break time-symmetry without
introducing  any approximations. The basic ideas of the approach, in detail, are examined in the works~\cite{liuv1},~\cite{ppt},~\cite{liuv2}~-~\cite{pp3}. Here, I briefly mention some specific features. 
Let me examine the equation which determines the density operator $\rho$ - Liouville-von Neumann equation 
\begin{equation}\label{ter}
i\frac{\partial\rho(t)}{\partial t}= L\rho(t).
\end{equation}
Liouvillian $L$ has the form
\begin{align}\label{liuvil}
L=H\times 1 - 1\times H,
\end{align}
where symbol "$\times$" denotes the operation $(A\times
B)\rho$=A$\rho$B, and $H$ is the total Hamiltonian, with $H=H_{0}+H_{I}$ being the sum of the free part $H_{0}$ and the interaction  part $H_{I}$. Similarly, the operator $L$ 
can be written down as the sum of the free part $L_{0}$, that depends
on the  Hamiltonian $H_{0}$ and the interaction part $L_{I}$,
which depends on $H_{I}$ : $L=L_{0} + L_{I}$.
Usually, for the operator $L$ the eigenvalues problem is formulated as follows
\begin{equation}\label{lv2}
L\mid f_{\alpha,\beta}\rangle\rangle =\tilde{w}_{\alpha,\beta}\mid f_{\alpha,\beta}\rangle\rangle,
\end{equation}
where  $\mid f_{\alpha,\beta}\rangle\rangle\equiv\mid\psi_{\alpha}><\psi_{\beta}\mid $, 
$\tilde{w}_{\alpha,\beta}=\tilde{E}_{\alpha}-\tilde{E}_{\beta}$ being the eigenstate and eigenvalue of Liouvillian $L$. The values $\mid\psi_{\alpha}>$ 
($\mid\psi_{\beta}>$),  $\tilde{E}_{\alpha}$ ($\tilde{E}_{\beta}$),
are the eigenstate and eigenvalue of the total Hamiltonian $H$, respectively (in this case
$H\mid\psi_{\alpha}>=\tilde{E}_{\alpha}\mid\psi_{\alpha}>$).
The equation~\eqref{lv2} can be rewritten by using the correlation index $\nu$
\begin{equation}\label{lv}
L\mid f_{\nu}\rangle\rangle=\tilde{w}^{\nu}\mid f_{\nu}\rangle\rangle,
\end{equation}
where $\tilde{w}^{\nu}\equiv \tilde{w}_{\alpha,\beta}$ and $\mid f_{\nu}\rangle\rangle\equiv \mid f_{\alpha,\beta}\rangle\rangle$.
In the equation~\eqref{lv}  $\nu = 0$ if $\alpha=\beta$
which is the set of diagonal operators $\mid\psi_{\alpha}><\psi_{\alpha}\mid$ or "vacuum of correlations" and $\nu\neq 0$ is the remaining off-diagonal case. The details of the theory of correlations can be found, for example, in the works~\cite{pp3},~\cite{ps}.
Eq.~\eqref{ter} is time reversible and does not describe the irreversible evolution.
Making the well known approximations from equation~\eqref{ter} it is possible to obtain the kinetic Pauli master equation; however, I will discuss an alternative approach based 
on the new formulation of the eigenvalues problem
\begin{align}\label{ff3u}
L\mid\Psi^{\nu}_{j}\rangle\rangle=Z^{\nu}_{j}\mid\Psi^{\nu}_{j}\rangle\rangle,~ \langle\langle
\tilde{\Psi}^{\nu}_{j}\mid L=\langle\langle \tilde{\Psi}^{\nu}_{j}\mid Z^{\nu}_{j},
\end{align}
where $Z^{\nu}_{j}$ are the complex values, and $j$ is a degeneracy index, since one type of correlation index can correspond to different states.
In this case, the complex eigenvalues $Z^{\nu}_{j}$ of the Liouville operator cannot be reduced to the simple difference of two eigenvalues of the Hamiltonian (as in the case~\eqref{lv2}) and eigenstates $\mid\Psi^{\nu}_{j}\rangle\rangle$, $\langle\langle \tilde{\Psi}^{\nu}_{j}\mid$  are not the multiplications of the wave functions (they cannot be obtained from the Schroedinger equation). 
As it was shown  by the Brussels-Austin group, this formulation has a broken time symmetry and leads to the rigorous description of irreversibility. \\ 
Further development of the approach is based on the theory of subdynamics.
Subdynamics is a term coined for the construction of a complete set of spectral projectors $\Pi^{\nu}$ 
\begin{align}\label{poper}
\Pi^{\nu}=\sum\limits_{j}|\Psi^{\nu}_{j}\rangle\rangle\langle\langle\widetilde{\Psi}^{\nu}_{j}|.
\end{align}
The projectors $\Pi^{\nu}$ satisfy the following relations
\begin{equation}\label{com}
\begin{split}
\Pi^{\nu}L=L\Pi^{\nu},~&(\text{commutativity});~
\sum\limits_{\nu}\Pi^{\nu}=1,~(\text{completeness});\\
&\Pi^{\nu}\Pi^{\nu'}=\Pi^{\nu}\delta_{\nu\nu'},~(\text{orthogonality}).
\end{split}
\end{equation}
Taking~\eqref{com}, it is possible to write down the density operator
$\rho$ in the form
\begin{align}\label{poper}
\rho=\sum\limits_{\nu}\Pi^{\nu}\rho=\sum\limits_{\nu}\rho^{\nu},
\end{align}
where $\rho^{\nu}\equiv\Pi^{\nu}\rho$.    
Because $\Pi^{\nu}L=L\Pi^{\nu}$, the operators $\rho^{\nu}$ satisfy the separate equations of motion~\cite{pass}
\begin{align}\label{poper2}
i\frac{\partial\Pi^{\nu}\rho(t)}{\partial t}= \Pi^{\nu}L\Pi^{\nu}\rho(t).
\end{align}
In other words, dynamical evolution is decomposed into independent "subdynamics"; it is specified by the single parameter $\nu$. 
Operator $\Pi^{\nu}$ is determined by the expression
\begin{align}\label{pca3}
\Pi^{\nu}=(P^{\nu}+C^{\nu})A^{\nu}(P^{\nu}+D^{\nu}),
\end{align}
where 
\begin{align}\label{pca2}
A^{\nu}=P^{\nu}(P^{\nu}+D^{\nu}C^{\nu})^{-1}P^{\nu}
\end{align}
and $C^{\nu}$, $D^{\nu}$ are kinetic operators, $P^{\nu}$ is a
projection operator. Operator
$C^{\nu}$ creates correlations other than the $\nu$ correlations, while
$D^{\nu}$ is destruction operator.  (operators $C^{\nu}$, $D^{\nu}$, $P^{\nu}$ can be found in the works~\cite{liuv1},~\cite{liuv2},~\cite{pp3} and in Appendix A).
As it has been shown~\cite{liuv1}, Eq.~\eqref{poper2} can be reduced to the Markovian kinetic equation       \begin{equation}\label{e110}
\begin{split}
i\frac{\partial P^{\nu}\rho^{\nu}(t)}{\partial t}=\vartheta^{\nu}_{C}P^{\nu}\rho^{\nu}(t).
\end{split}
\end{equation}
Here the value $\vartheta^{\nu}_{C}$ is a non-Hermitian dissipative operator which plays the main role in the nonequalibrium dynamics  
\begin{align}\label{ae9}
\begin{split}
&\vartheta^{\nu}_{C}= P^{\nu}L(P^{\nu} + C^{\nu}).
\end{split}
\end{align}
As it was shown in the work~\cite{shirm2}, the density operator $\rho$ can be determined by the kinetic operators 
$C^{\nu}$, $D^{\nu}$, $\vartheta^{\nu}_{C}$
\begin{equation}\label{pro}
\begin{split}
P^{0}\rho(t)=P^{0}\exp(-i\vartheta^{0}_{C}t)A^{0}P^{0}\rho(0)+
\sum\limits_{\nu\neq 0}P^{0}C^{\nu}\exp(-i\vartheta^{\nu}_{C}t)A^{\nu}D^{\nu}
P^{0}\rho(0).
\end{split}
\end{equation}
The Eq.~\eqref{pro} contains the Markovian part in the first term and non-Markovian contribution in the second term. The expression~\eqref{pro} for the density operator is essential for our further analysis. \\
In the paper I examine the two-nucleus spin systems which consist of a pair of uniform nuclei with a spin of $1/2$.
The spin systems are determined by the H(1) and C(13) nuclei. The interaction between the magnetic moments of the  nuclei means that the pair of nuclei behave as  connected system. For example, the transition of one nucleus from the upper level to the lower (the effect of the Nuclear Magnetic Resonance (NMR)) can cause the energy transition of the adjacent nucleus. In the work the energy transitions are considered to be irreversible processes. The two-nucleus spin NMR  systems are analyzed in the framework of the subdynamics theory. In Section 2 the theory of magnetic interaction of the nuclei is discussed. The solution to the complex eigenvalues problem is given in Section 3. The time irreversible evolution of the two-nucleus spin systems is investigated in Section 4. In Section 5, the results of the work, and quantum Zeno effect are discussed.

\section{The magnetic interaction}
Let me examine the pair of the uniform nuclei which are located in the external magnetic field, directed along z axis - $H_{z}$. The Hamiltonian of this system can be written down in the form 
\begin{equation}\label{e2}
\begin{split}
&{{H}}_{IH_{z}+NN}=-\gamma_{N}H_{z}({I}_{z1}+{I}_{z2})-\gamma_{N}^{2}r^{-3}[
3(\vec{I}_{1}\vec{n})(\vec{I}_{2}\vec{n})-\vec{I}_{1}\vec{I}_{2}]\\
&\equiv{{H}}_{IH_{z}}+H_{NN}.
\end{split}
\end{equation} 
Here the first term, ${{H}}_{IH_{z}}$, corresponds to the interaction of the system with the magnetic field $H_{z}$ and the second, $H_{NN}$, is the interaction between the magnetic moments of the nuclei.
The values $\vec{I}_{1}$, $\vec{I}_{2}$  are the spin vectors of the first and the second  nucleus, ${I}_{z1}$, ${I}_{z2}$ are their z - components, $r$ is the distance between the nuclei, $\vec{n}$ - unit vector in the direction from the first nucleus to the second, 
$\gamma_{N}$ is a gyromagnetic ratio $\gamma_{N}=\mu/I$, 
where $\mu$, $I$ are the magnetic moment and the spin of the nucleus. \\ 
Further examination of the magnetic interaction of the nuclei will be  based on the works~\cite{slonim},~\cite{Corio}.  
At the beginning let, the interaction between the nuclei be absent. In this case, the projections of the summary spin of the system $I_{z}={I}_{z1}+{I}_{z2}$ on z axis have four different values: 1,~0,~0,~-1.
The projections correspond to the four different states: the spins are directed along the field $H_{z}$ ($\uparrow\uparrow$); they are directed differently ($\downarrow\uparrow$) or ($\uparrow\downarrow$); and the spins are directed opposite to the field ($\downarrow\downarrow$). To describe these four different states, the four eigenstates can be determined
\begin{equation}\label{e4}
\begin{split}
&\psi_{1}=\begin{pmatrix}1&\\ 0\end{pmatrix}\begin{pmatrix}1&\\ 0\end{pmatrix}=(\alpha\alpha), ~~~~\psi_{2}=\begin{pmatrix}0&\\ 1\end{pmatrix}\begin{pmatrix}1&\\ 0\end{pmatrix}=(\beta\alpha),\\
&\psi_{3}=\begin{pmatrix}1&\\ 0\end{pmatrix}\begin{pmatrix}0&\\ 1\end{pmatrix}=(\alpha\beta), ~~~~\psi_{4}=\begin{pmatrix}0&\\ 1\end{pmatrix}\begin{pmatrix}0&\\
1\end{pmatrix}=(\beta\beta),
\end{split}
\end{equation}
where the function $\alpha$ corresponds to the spin state ($\uparrow$)  (the projection of the spin of the single nucleus on z axis is $1/2$) and the function $\beta$ corresponds to the  spin state ($\downarrow$) (the projection of the spin on z axis is $-1/2$).\\
Let me note the validity of the following expressions (where the units $\hbar =1$ and the speed of light $c=1$ are used ): \\
1.~The action of the spin operator on the functions ($i$=1 or 2)
\begin{equation}\label{e5}
\begin{split}
&I_{xi}~\alpha=\frac{1}{2}\beta,~~~~I_{yi}~\alpha=\frac{i}{2}\beta,~~~~I_{zi}~\alpha=\frac{1}{2}\alpha,\\
&I_{xi}~\beta=\frac{1}{2}\alpha,~~~~I_{yi}~\beta=-\frac{i}{2}\alpha,~~~~I_{zi}~\beta=-\frac{1}{2}\beta;
\end{split}
\end{equation}
\begin{equation}\label{e6}
\begin{split}
I_{i}^{2}~\alpha=\frac{3}{4}\alpha,~~~~I_{i}^{2}~\beta=\frac{3}{4}\beta.
\end{split}
\end{equation}  
2. The multiplication of the pairs of identical functions
\begin{equation}\label{e7}
\begin{split}
(\alpha\alpha)(\alpha\alpha) &=(\alpha\beta)(\alpha\beta)=(\beta\alpha)(\beta\alpha)=(\beta\beta)(\beta\beta)=1
\end{split}
\end{equation}
and different functions
\begin{equation}\label{e77}
\begin{split}
&(\alpha\alpha)(\alpha\beta)=0,~~~~(\alpha\alpha)(\beta\alpha)=0~~\text {and so on}. 
\end{split}
\end{equation}
Furthermore, 
\begin{equation}\label{e809}
\begin{split}
&I_{z}(\alpha\alpha)=(\alpha\alpha),~~~~I_{z}(\beta\alpha)=0(\beta\alpha),\\
&I_{z}(\alpha\beta)=0(\alpha\beta),~~~~I_{z}(\beta\beta)=-(\beta\beta).
\end{split}
\end{equation}
Thus, from the expressions~\eqref{e809} it follows that the functions ($\alpha\alpha$), 
($\beta\alpha$), ($\alpha\beta$), ($\beta\beta$) correspond to the spin states ($\uparrow\uparrow$), ($\downarrow\uparrow$), ($\uparrow\downarrow$), ($\downarrow\downarrow$), respectively.
However, only the functions ($\alpha\alpha$), ($\beta\beta$) are the eigenstates of the summary operator $I^{2}=(\vec{I}_{1}+\vec{I}_{2})^{2}$
\begin{equation}\label{e10}
\begin{split}
&I^{2}(\alpha\alpha)=2(\alpha\alpha),~~~~I^{2}(\beta\beta)=2(\beta\beta).
\end{split}
\end{equation}
The functions ($\alpha\beta$), ($\beta\alpha$) are not the same
\begin{equation}\label{e100}
\begin{split}
I^{2}(\alpha\beta)=((\alpha\beta)+(\beta\alpha)),~~~~
I^{2}(\beta\alpha)=((\alpha\beta)+(\beta\alpha)).
\end{split}
\end{equation}
The results~\eqref{e100} are not satisfactory for the description of the nuclei spin system. However, if we substitute the functions $\frac{1}{\sqrt{2}}((\alpha\beta)+(\beta\alpha))$ and $\frac{1}{\sqrt{2}}((\alpha\beta)-(\beta\alpha))$ for those of ($\alpha\beta$) and ($\beta\alpha$), then the four new functions are defined as 
\begin{equation}\label{e11}
\begin{split}
\phi_{1}=(\alpha\alpha),~~~~\phi_{2}&=\frac{1}{\sqrt{2}}((\alpha\beta)+(\beta\alpha)),~~~~\phi_{3}=(\beta\beta),~~~~\\
&\phi_{4}=\frac{1}{\sqrt{2}}((\alpha\beta)-(\beta\alpha))
\end{split}
\end{equation}
will be the eigenstates of the operators $I_{z}$, $I^{2}$ with the eigenvalues 1, 0, -1, 0  and 2, 2, 2, 0, respectively. 
From the aforesaid it follows that the pair of the noninteracting spins 1/2 in the states $\phi_{1}$, $\phi_{2}$, $\phi_{3}$ behaves as one particle with the spin 1, while in the state $\phi_{4}$ as the particle with the spin 0. Since the functions $\phi_{1}$, $\phi_{2}$, $\phi_{3}$, $\phi_{4}$ are the eigenstates of the operators $I_{z}$, $I^{2}$, it is convenient use them to describe the interacting spin system.\\ 
Now our aim is to determine the energy of the interacting two-nucleus spin system placed into the external magnetic field $H_{z}$. The solution to this problem can be conveniently carried out in the matrix representation of the Hamiltonian~\eqref{e2}. 
Calculating the matrix elements of the operator ${{H}}_{IH_{z}}$ with the use of the functions~\eqref{e11} (in according to the rules~\eqref{e5}~-~\eqref{e77}) leads to 
\begin{equation}\label{e134}
\begin{split}
{{H}}_{IH_{z}}=-\gamma_{N}H_{z}\begin{pmatrix}1 & 0 & 0 & 0 \\ 0 & 0 & 0 & 0\\0 & 0 & -1 & 0\\0 & 0& 0 & 0\end{pmatrix}.
\end{split}
\end{equation}
Let me examine the operator $H_{NN}$. In the Cartesian coordinates the spin operators  $\vec{I}_{1}$, $\vec{I}_{2}$ and the unit vector $\vec{n}$ can be represented as
\begin{equation}\label{e142}
\begin{split}
&\vec{I}_{i}=\vec{i}I_{xi}+\vec{j}I_{yi}+\vec{k}I_{zi},\\
&\vec{n}=\vec{i}\cos(\alpha)+\vec{j}\cos(\beta)+\vec{k}\cos(\gamma),
\end{split}
\end{equation} 
here $\vec{i}$, $\vec{j}$, $\vec{k}$ are the unit vectors directed along the coordinate axes and $\alpha$, $\beta$, $\gamma$ angles between the direction from one nucleus to another and the coordinate axes $x$, $y$, $z$. Let me switch over to polar coordinate system. In this case,    
\begin{equation}\label{e152}
\begin{split}
\cos(\alpha)=\sin(\theta)\cos(\varphi),~~\cos(\beta)=\sin(\theta)\sin(\varphi),~~\cos(\gamma)=\cos(\theta),
\end{split}
\end{equation}
with $\theta$ and $\varphi$ being polar and azimuthal angles. Using the expressions~\eqref{e142},~\eqref{e152}
we obtain
\begin{equation}\label{e17}
\begin{split}
H_{NN}=&[I_{z1}I_{z2}-\frac{1}{4}(I_{1}^{-}I_{2}^{+}+I_{1}^{+}I_{2}^{-})]\gamma_{N}^{2}Y_{0}
-\frac{3}{2}(I_{1}^{+}I_{z2}+I_{z1}I_{2}^{+})\gamma_{N}^{2}Y_{1}\\
&-\frac{3}{2}(I_{1}^{-}I_{z2}+
I_{z1}I_{2}^{-})\gamma_{N}^{2}Y_{1}^{*}
-\frac{3}{4}I_{1}^{+}I_{2}^{+}\gamma_{N}^{2}Y_{2}-\frac{3}{4}I_{1}^{-}I_{2}^{-}\gamma_{N}^{2}Y_{2}^{*},
\end{split}
\end{equation}
where the action "*" is the complex conjugation,
\begin{equation}\label{e189}
\begin{split}
I_{i}^{\pm}=(I_{x1}\pm iI_{yi})
\end{split}
\end{equation}
and
\begin{equation}\label{e193}
\begin{split}
&Y_{0}=r^{-3}(1-3\cos^{2}(\theta)),\\
&Y_{1}=r^{-3}\sin(\theta)\cos(\theta)\exp(-i\varphi),\\
&Y_{2}=r^{-3}\sin^{2}(\theta)\exp(-2i\varphi).
\end{split}
\end{equation}
Calculating the matrix elements of the operator $H_{NN}$ leads to the matrix representation 
\begin{equation}\label{e2023}
\begin{split}
H_{NN}=\frac{1}{4}\gamma_{N}^{2}\begin{pmatrix}Y_{0} & -3\sqrt{2}Y_{1} & -3Y_{2} & 0 \\ 
-3\sqrt{2}Y_{1}^{*} & -2Y_{0} & 3\sqrt{2}Y_{1} & 0\\-3Y_{2}^{*} & 3\sqrt{2}Y_{1}^{*} & Y_{0} & 0
\\0 & 0& 0 & 0\end{pmatrix}.
\end{split}
\end{equation}
It has been shown~\cite{slonim} that the influence of the nondiagonal elements in the first approximation can be disregarded since the changes of the energy levels, caused by them, 
are small. Taking into account the latter fact and the expressions~\eqref{e2},~\eqref{e134},~\eqref{e2023} for the Hamiltonian ${{H}}_{IH_{z}+NN}$, we have 
\begin{equation}\label{e245}
\begin{split}
{{H}}_{IH_{z}+NN}=-\gamma_{N}\begin{pmatrix}H_{z}-\frac{1}{4}\gamma_{N} Y_{0} & 0 & 0 & 0 \\ 
0 & \frac{1}{2}\gamma_{N} Y_{0} & 0 & 0\\0 & 0 & -H_{z}-\frac{1}{4}\gamma_{N} Y_{0} & 0
\\0 & 0& 0 & 0\end{pmatrix}.
\end{split}
\end{equation}
It follows from the expression~\eqref{e245} that the interaction of the nuclear spin system with the external stationary magnetic field $H_{z}$ as well as interaction between the magnetic moments of the nuclei lead to the splitting of the system energy levels into the values 
\begin{equation}\label{e276}
\begin{split}
&E_{1}=-\gamma_{N} H_{z}+\frac{1}{4}\gamma_{N}^{2} Y_{0},~~E_{2}=-\frac{1}{2}\gamma_{N}^{2} Y_{0},\\
&E_{3}=\gamma_{N} H_{z}+\frac{1}{4}\gamma_{N}^{2} Y_{0},~~E_{4}=0.
\end{split}
\end{equation}
From the expression~\eqref{e245} it  follows that the singlet state $\phi_{4}$ is not affected by the interaction since it corresponds to zero summary spin and consequently to zero magnetic moment. \\
The transition  between the energy levels~\eqref{e276} can occur as a result of the external, variable magnetic field 
$\vec{H}_{1}$ perpendicular to the field $H_{z}$ (the NMR effect) influencing the system.
In the simplest case, the components of the field $\vec{H}_{1}$ change according to the law
\begin{equation}\label{e229} 
H_{x}=H_{1}\cos(\omega t),~~H_{y}=-H_{1}\sin(\omega t),
\end{equation}
where $\omega$ is the frequency of the rotating field in the $xy$ plane.\\
In this case, the Hamiltonian of the disturbance  $H_{IH_{xy}}$  can be written down as 
\begin{equation}\label{e284}
\begin{split}
&H_{IH_{xy}}=-\gamma_{N}H_{1}(I_{x}\cos(\omega t)-I_{y}\sin(\omega t)),
\end{split}
\end{equation}
where $I_{x}=I_{x1}+I_{x2}$, $I_{y}=I_{y1}+I_{y2}$. 
The expressions~\eqref{e284} and~\eqref{e11} lead to the matrix representation of $H_{IH_{xy}}$ 
\begin{equation}\label{e30}
\begin{split}
H_{IH_{xy}}=-\frac{\gamma_{N}H_{1}}{\sqrt{2}}
\begin{pmatrix}0 & \exp(i\omega t) & 0 & 0 \\ 
\exp(-i\omega t) & 0 & \exp(i\omega t) & 0\\0 &\exp(-i\omega t) & 0 & 0
\\0 & 0& 0 & 0\end{pmatrix}.
\end{split}
\end{equation}
Since the transitions are caused by the variable disturbance $H_{IH_{xy}}$, then it is obvious from the expression~\eqref{e30} that the ones are permitted only between the states $\phi_{1}$,  $\phi_{2}$ and between the states $\phi_{2}$, $\phi_{3}$. In the first case, the transition energy $\Delta E_{12}$ is determined by
\begin{equation}\label{e253}
\begin{split}
\Delta E_{12}=\gamma_{N}\Bigl{(}H_{z}+\frac{3}{4}\gamma_{N} r^{-3}(3\cos^{2}(\theta)-1)\Bigr{)},
\end{split}
\end{equation}
in the second case, the transition energy $\Delta E_{23}$ is determined as follows
\begin{equation}\label{e2564}
\begin{split}
\Delta E_{23}=\gamma_{N}\Bigl{(}H_{z}-\frac{3}{4}\gamma_{N} r^{-3}(3\cos^{2}(\theta)-1)\Bigr{)}.
\end{split}
\end{equation}
The expression~\eqref{e30} corresponds to the monochromatic model. In order to pass to the general case it is necessary to make the following replacements
\begin{equation}\label{e74}
\begin{split}
&H_{1}\exp(i\omega t) \rightarrow \int\limits_{-\infty}^{\infty}A(\omega)H_{1}\exp(i\omega t)d\omega ,\\
&H_{1}\exp(-i\omega t) \rightarrow \int\limits_{-\infty}^{\infty}B(\omega)H_{1}\exp(-i\omega t)d\omega,
\end{split}
\end{equation}
where the coefficients $A(\omega)$, $B(\omega)$ determine the distribution of the radiated energies. \\
As it has been shown, in the first approximation~\eqref{e245} the functions~\eqref{e11} can be defined as the eagenstates of the Hamiltonian~\eqref{e2}.
Therefore, I assume that any function which describes the behavior of our spin system can be represented as expansion in terms of the latter 
\begin{equation}\label{e321}
\begin{split}
&\phi=a_{1}\phi_{1}\exp(-iE_{1}t)+a_{2}\phi_{2}\exp(-iE_{2}t)+a_{3}\phi_{3}\exp(-iE_{3}t)\\
&+a_{4}\phi_{4}\exp(-iE_{4}t),
\end{split}
\end{equation}
where the energies $E_{1}$, $E_{2}$, $E_{3}$, $E_{4}$ satisfy the relationships~\eqref{e276}.\\ 
For the hermitian conjugated function I have
\begin{equation}\label{e331} 
\begin{split}
&\phi^{\dagger}=a_{1}^{*}\phi_{1}^{\dagger}\exp(iE_{1}t)+a_{2}^{*}\phi_{2}^{\dagger}\exp(iE_{2}t)+a_{3}^{*}\phi_{3}^{\dagger}\exp(iE_{3}t)\\
&+a_{4}^{*}\phi_{4}^{\dagger}\exp(iE_{4}t).
\end{split}
\end{equation}
The expressions which were obtained in this section are the basis for the further model development. 
 
\section{Complex eigenvalues problem}
Let me examine the transition from the level $E_{3}$ (the projection of the spin is -1) to the level $E_{2}$ (the projection of the spin is 0) as an irreversible process, when the energy $\Delta E_{23}$ is emitted. The solution to this problem  will be carried out on the basis of second quantization method. 
In this case, the functions $A(\omega)$, $B(\omega)$ are defined as operators 
\begin{equation}\label{e667}
\begin{split}
A(\omega)\rightarrow g(\omega)a^{\dagger}(\omega),~~B(\omega)\rightarrow g(\omega)a(\omega).
\end{split}
\end{equation}
Here $a^{\dagger}(\omega)$, $a(\omega)$ are the operators of creation and destruction of  a photon with the energy $\omega$.
I assume, these operators satisfy the commutation relation
\begin{equation}\label{e8}
\begin{split}
[a(\omega),a^{\dagger}(\omega ')]=\delta(\omega - \omega ').
\end{split}
\end{equation}
The value $g(\omega)$ is determined by the fact that its  square  is the distribution function
satisfying the condition 
\begin{equation}\label{e9}
\begin{split}
\int\limits_{-\infty}^{\infty}g^{2}(\omega)d\omega =1.
\end{split}
\end{equation}
The functions~\eqref{e321},~\eqref{e331} are defined  now as operators too.
The coefficients $a_{i}$ in the expression~\eqref{e321}, where $i=1,2,3,4$ specify destruction operators of the state with  projections of the spin 1,0,-1,0, respectively. 
For $a_{i}^{*}$ in~\eqref{e331} we must make
$a_{i}^{*}\rightarrow a_{i}^{\dagger}$.
The values $a_{i}^{\dagger}$ specify creation operators.
It is assumed that the operators $a_{i}$, $a_{i}^{\dagger}$ satisfy the commutation relation 
\begin{equation}\label{e34}
\begin{split}
[a_{m},a_{m'}^{\dagger}] = \delta_{mm'},~~m~(m')=\pm1,~0.
\end{split}
\end{equation}
The Hamiltonian of the two-nucleus spin NMR system $H_{NM}$ have to be determined as
\begin{equation}\label{e2534}
H_{NM}=\phi^{\dagger}({{H}}_{IH_{z}+NN}+H_{IH_{xy}})\phi \equiv {{H}}_{IH_{z}}+H_{NN}+ H_{IH_{xy}},
\end{equation}
where the operators in the brackets  are determined by the expressions~\eqref{e2},~\eqref{e284} (in the expression~\eqref{e2534} the previous designations of the Hamiltonians are preserved). \\
In the model, the complete Hamiltonian of the system I define as $H=H_{0}+H_{I}$, where 
$H_{0}$ is a free part $H_{0}=H_{\omega}+{H}_{IH_{z}+NN}$ and $H_{I}=H_{IH_{xy}}\equiv \lambda V$ is a interaction part (with $\lambda$ being a coupling constant).  The free photon Hamiltonian $H_{\omega}$ is determined as follows 
\begin{equation}\label{e665po}
\begin{split}
H_{\omega}=\int\limits_{-\infty}^{\infty}\omega a^{\dagger}(\omega)a(\omega)d\omega.
\end{split}
\end{equation}
In this case, $H_{\omega}\mid\omega>=\omega\mid\omega>$, where the state $\mid\omega>=a^{\dagger}(\omega)\mid vac>$  is a eigenstate of the free Hamiltonian $H_{\omega}$ and $\omega$ is a photon energy ( $\mid vac>$ is a vacuum of the state: $a(\omega)\mid vac>=0$). 
I determine the states with projection of the spin on z axis equal $\pm 1$ and $0$: $\mid\pm 1> =a_{1,3}^{\dagger}\mid vac>$,  $\mid 0> =a_{2,4}^{\dagger}\mid vac>$, respectively. The states $\mid\pm 1>$, $\mid 0>$ are the eigenstates of the Hamiltonian ${{H}}_{IH_{z}+NN}$:
${{H}}_{IH_{z}+NN}\mid\pm 1>=E_{1,3}\mid\pm 1>$ and ${{H}}_{IH_{z}+NN}\mid 0>=E_{2,4}\mid 0>$.
Furthermore, another two states  $\mid \pm 1, \omega>$, $\mid 0, \omega>$ correspond to the nucleus - photon system. For them, the relationships $H_{0}\mid \pm 1, \omega>=(\omega + E_{1,3})\mid \pm1, \omega>$, $H_{0}\mid 0, \omega>=(\omega + E_{2,4})\mid 0, \omega>$ are valid, where  $\mid \pm1, \omega>=a_{1,3}^{\dagger}a^{\dagger}(\omega)\mid vac>$ and  $\mid 0, \omega>=a_{2,4}^{\dagger}a^{\dagger}(\omega)\mid vac>$.\\
I determine the eigenvalues problem for the complete Hamiltonian $H=H_{\omega}+{H}_{IH_{z}+NN}+H_{IH_{xy}}$ for the state that corresponds to the projection of the spin on z axis equal to $- 1$.
As it has been mentioned above (Introduction: the relationships ~\eqref{ht1},~\eqref{gamma}), the description of the transition $E_{3}\rightarrow E_{2}$ (transition $\mid- 1>\rightarrow\mid 0,\omega>$), as an irreversible process, has to be executed outside the Hilbert space. 
It is assumed that the eigenvalue $Z_{-1}$ of the
Hamiltonian is complex. Then I have the equations for the right eigenstate $\mid\varphi_{-1}>$ and
for the left eigenstate $<\widetilde{\varphi}_{-1}\mid$ of the Hamiltonian $H$  
\begin{equation}\label{e165}
\begin{split}
H\mid\varphi_{-1}>=Z_{-1}\mid\varphi_{-1}>,
~<\widetilde{\varphi}_{-1}\mid H = <\widetilde{\varphi}_{-1}\mid
Z_{-1}.
\end{split}
\end{equation}
According to the method developed in the works~\cite{liuv1} and~\cite{ppt}, I expand the values $\mid\varphi_{-1}>$,
$<\widetilde{\varphi}_{-1}\mid$, $Z_{-1}$ in the
series
\begin{align}\label{ht3}
&\mid\varphi_{-1}>=\sum\limits_{n=0}^{\infty}\lambda^{n}\mid\varphi_{-1}^{(n)}>,~
<\widetilde{\varphi}_{-1}\mid=\sum\limits_{n=0}^{\infty}\lambda^{n}<\widetilde{\varphi}_{-1}^{(n)}\mid, \\
&Z_{-1}=\sum\limits_{n=0}^{\infty}\lambda^{n}Z_{-1}^{(n)},
\end{align}
where
\begin{align}\label{e412}
\begin{split}
&\mid\varphi_{-1}^{(0)}>=\mid -1>,~
<\widetilde{\varphi}_{-1}^{(0)}\mid=<-1 \mid,~
Z_{-1}^{(0)}=E_{3},\\
&~\lambda \equiv e - \text{the elementary charge.}
\end{split}
\end{align}
The complex eigenvalue problem~\eqref{e165} is solved analogously to the works~\cite{ppt},~\cite{shirm2}. 
In this case, for the coefficients $Z_{-1}^{(n)}$ it is easy to obtain
\begin{equation}\label{e586}
\begin{split}
Z_{-1}^{(n)} =<-1\mid V\mid\varphi_{-1}^{(n-1)}> - \sum\limits_{l=1}^{n-1}Z_{-1}^{(l)}<-1\mid\varphi_{-1}^{(n-l)}>,
\end{split}
\end{equation}
where the interaction term is determined by the expression
\begin{equation}\label{e346}
\begin{split}
&<-1\mid V\mid\varphi_{-1}^{(n-1)}> = -\frac{\gamma_{N}H_{1}}{e\sqrt{2}}\int\limits_{-\infty}^{\infty} g(\omega)\exp{(-i(E_{2}-E_{3}+\omega)t)} \\
&\times< 0,\omega\mid\varphi_{-1}^{(n-1)}> d\omega.
\end{split}
\end{equation}
Using~\eqref{e346}, for $Z_{-1}$ the following result is valid
\begin{equation}\label{e787}
\begin{split}
&Z_{-1} - E_{3} =  -\frac{\gamma_{N}H_{1}}{\sqrt{2}}\int\limits_{-\infty}^{\infty}  g(\omega)\exp{(-i(E_{2}-E_{3}+\omega)t)} \\
&\times < 0,\omega\mid\varphi_{-1}> d\omega -
(Z_{-1} - E_{3})(<-1\mid\varphi_{-1}>-1).
\end{split}
\end{equation}
The determination of the matrix element $< 0,\omega\mid\varphi_{-1}>$ follows from the obvious relationship
\begin{equation}\label{e890}
\begin{split}
&< 0, \omega\mid\Bigr{(}H_{0}\sum\limits_{n=0}^{\infty}e^{n}\mid\varphi_{-1}^{(n)}> + e V \sum\limits_{n=0}^{\infty}e^{n}\mid\varphi_{-1}^{(n)}>\Bigr{)} \\
&=< 0, \omega\mid\sum\limits_{n=0}^{\infty}e^{n}Z_{-1}^{(n)}\sum\limits_{n'=0}^{\infty}e^{n'}\mid\varphi_{-1}^{(n')}>,
\end{split}
\end{equation} 
hence it follows 
\begin{equation}\label{e923}
\begin{split}
&< 0,\omega\mid\varphi_{-1}^{(n)}> =  \frac{-1}{E_{2}+\omega - E_{3}}\\
&\times\Bigl{(}< 0,\omega\mid V\mid\varphi_{-1}^{(n-1)}> - \sum\limits_{l=1}^{n}Z_{-1}^{(l)}< 0,\omega\mid\varphi_{-1}^{(n-l)}> \Bigr{)}.
\end{split}
\end{equation}
The transition  $\mid- 1>\rightarrow\mid 0,\omega>$ is a time ordered irreversible process.
Therefore, the time ordering of the expression~\eqref{e923} has to be introduced. For this purpose, it is necessary to make the replacement 
\begin{equation}\label{e1009}
\begin{split}
\frac{1}{E_{2}+\omega - E_{3}}\rightarrow\frac{1}{E_{2}+\omega - E_{3}-i\varepsilon_{\mu\nu}}.
\end{split}
\end{equation}
Here $\nu$ - correlation index corresponds to the initial state $\mid - 1>$ and $\mu$ - correlation index corresponds to the final state $\mid 0,\omega>$. In our case, it is possible to introduce the redesignation $\varepsilon_{\mu\nu}\equiv\varepsilon >0$ ( the real infinitesimal $\varepsilon_{\mu\nu}$ determination details can be found in the work~\cite{ppt}).
For the first matrix element of the expression~\eqref{e923} I have
\begin{equation}\label{e119}
\begin{split}
&< 0,\omega\mid V\mid\varphi_{-1}^{(n-1)}> = -\frac{\gamma_{N}H_{1}}{e\sqrt{2}}g(\omega)
\exp(i(E_{2}-E_{3}+\omega)t) \\
&\times<-1\mid\varphi_{-1}^{(n-1)}>.
\end{split}
\end{equation}
Substituting the expressions~\eqref{e1009},~\eqref{e119} for~\eqref{e923} and summing up $\sum\limits_{n=1}^{\infty}e^{n}$ result in
\begin{equation}\label{e120}
\begin{split}
&< 0,\omega\mid\varphi_{-1}> = \frac{\gamma_{N}H_{1}}{\sqrt{2}}g(\omega) \exp(i(E_{2}-E_{3}+\omega)t) \\
&\times<-1\mid\varphi_{-1}>\frac{1}{E_{2}+\omega - Z_{-1} - i\varepsilon} .
\end{split}
\end{equation}
The expression~\eqref{e120} leads to 
\begin{equation}\label{e1334}
\begin{split}
Z_{-1} = E_{3} - \frac{\gamma_{N}^{2}H_{1}^{2}}{2}\int\limits_{-\infty}^{\infty} g^{2}(\omega)\frac{1}{E_{2}+\omega - Z_{-1} -i\varepsilon} d\omega .
\end{split}
\end{equation}
The formal expression
\begin{equation}\label{e143}
\begin{split}
\frac{1}{E_{2}+\omega - Z_{-1} -i\varepsilon}  \approx \textbf{P}\frac{1}{E_{2}+\omega -E_{3} } + i\pi\delta (E_{2}+\omega - E_{3}) 
\end{split}
\end{equation}
makes it possible  to rewrite the relationship for $Z_{-1}$ in the form 
\begin{equation}\label{e153}
\begin{split}
Z_{-1} = \bar{E}_{3} -i\gamma_{-1},
\end{split}
\end{equation}
where $\bar{E}_{3}$ is a renormalized energy of the initial state
\begin{equation}\label{e1690}
\begin{split}
&\bar{E}_{3} = E_{3} - \frac{\gamma_{N}^{2}H_{1}^{2}}{2}~\textbf{P}\int\limits_{-\infty}^{\infty} g^{2}(\omega)\frac{1}{E_{2}+\omega - E_{3}} d\omega, \\
&\text{\textbf{P} being the principal part  of the integral}
\end{split}
\end{equation}
and 
\begin{equation}\label{e1709}
\begin{split}
\gamma_{-1} = \frac{1}{2}\pi\gamma_{N}^{2}H_{1}^{2}g^{2}(\Delta E_{23}) .
\end{split}
\end{equation}
It is known~\cite{abragam} that the rate of the transition between two states $\mid m>$ and $\mid m'>$, where $m$ ($m'$) is the projection of the spin $\vec{I}$ on the z axis is defined by the expression
\begin{equation}\label{e1832}
\begin{split}
W_{m\rightarrow m'}=\frac{\pi}{2}\gamma_{N}^2H_{1}^{2}(I+m+1)(I-m)f(\omega),
\end{split}
\end{equation}
where $m'=m\pm1$, and $f(\omega)$ is a distribution function with the width $\Delta$.
The  distribution function describes, for example, Lorentzian or Gaussian distribution of energy. Lorentzian distribution is described by the expression
\begin{equation}\label{e867}
\begin{split}
f(\omega)=\frac{\delta}{\pi}\frac{1}{\delta^{2}+(\omega -\omega_{0})^{2}},~\delta=\frac{1}{2}\Delta
\end{split}
\end{equation}
with the norm
\begin{equation}\label{norm}
\int\limits_{-\infty}^{\infty}f(\omega)d\omega = 1.
\end{equation}
Since the singlet state $\phi_{4}$ corresponds to zero magnetic moment and does not participate in the energy transitions, we can conditionally assume the spin of the two-nucleus system participating in the interaction, $I=1$. In our case, $m=-1$ while $m'=0$. Then from the  expression~\eqref{e1832} it  follows that
\begin{equation}\label{f8865}
\begin{split}
W_{-1\rightarrow 0}=\pi\gamma_{N}^{2}H_{1}^{2}f(\omega).
\end{split}
\end{equation}
Comparing the results ~\eqref{e1709},~\eqref{f8865} we see  
\begin{equation}\label{e1834}
\begin{split}
W_{-1\rightarrow 0} /_{\omega = \omega_{0}} =2 \gamma_{-1}/_{g^{2}(\omega_{0})= f(\omega_{0})},
\end{split}
\end{equation}
where the energy $\omega_{0}\equiv\Delta E_{23}$ and  $g^{2}(\omega_{0})= f(\omega_{0})$. Thus, the expression obtained for $2 \gamma_{-1}$ coincides with the rate~\eqref{f8865} under the resonance condition $\omega =\omega_{0}$. The conclusion~\eqref{e1834} is in agreement with the results of the work~\cite{shirm2}, in which the transition $\mid-1/2>\rightarrow \mid1/2,\omega>$ was investigated. It has been shown that the expression for $2 \gamma_{-1/2}$ coincides with the rate $W_{-1/2\rightarrow 1/2}$ under the resonance condition $\omega_{0}=E_{-1/2}-E_{1/2}$. In the work~\cite{shirm}, where the irreversible evolution of the unstable $\pi^{-}$~-~meson was examined, the value $2\gamma_{\mathbf{p}_{\pi}}$ was defined as the rate of the $\pi^{-}$~-~meson decay.\\
Thus, on the basis of the Brussels-Austin group approach, the value of the rate is obtained as the solution to the eigenvalues problem, on the basis of the complex spectral representation without any mnemonic rules. 

\section{Evolution of the two-nucleus spin system}
The description of the irreversible evolution of the two-nucleus spin system is carried out on the basis of the subdynamics theory. I examine the matrix element \\$\langle \langle-1;-1\mid P^{0}\rho(t)\rangle \rangle$ (the expression~\eqref{pro}):
\begin{equation}\label{e1430}
\begin{split}
\langle \langle -1 ; -1\mid P^{(0)}\rho(t)\rangle \rangle &= \langle \langle -1; -1\mid \rho(t)\rangle \rangle\\ 
&\equiv  
< -1\mid \rho (t)\mid -1>,
\end{split}
\end{equation}
where the projection operator $P^{0}$ is determinated in Appendix A.
Let me represent the operator $A^{0}$~\eqref{pca2} as series expansion. 
In contrast to the work~\cite{shirm2}, where the opreator $A^{0}$ was examined in the first approximation, here I define it, taking into account the second term 
\begin{equation}\label{e2324}
\begin{split}
A^{0}=P^{0}(P^{0}+D^{0}C^{0})^{-1}P^{0}\approx P^{0} - D^{0}C^{0}.
\end{split}
\end{equation}
In this case, for the matrix element~\eqref{e1430} it is obvious  
\begin{equation}\label{e39853}
\begin{split}
\langle \langle -1;-1\mid\rho(t)\rangle \rangle &=\langle \langle -1;-1\mid\exp(-i\vartheta^{0}_{C}t)P^{0}\rho(0)\rangle \rangle \\
&- 
\langle \langle -1;-1\mid\exp(-i\vartheta^{0}_{C}t)D^{0}C^{0}P^{0}\rho(0)\rangle \rangle \\
&+\sum\limits_{\nu\neq 0}\langle \langle -1; -1\mid C^{\nu}\exp(-i\vartheta^{\nu}_{C}t)P^{\nu}D^{\nu}P^{0}\rho(0)\rangle \rangle. 
\end{split}
\end{equation} 
For the first term of the expression~\eqref{e39853} I have
\begin{equation}\label{e590}
\begin{split}
\langle \langle -1;-1\mid &\exp(-i\vartheta^{0}_{C}t)P^{0}\rho(0)\rangle \rangle \\
&=\langle \langle -1;-1\mid\exp(-i\vartheta^{0}_{C}t)\mid-1;-1\rangle \rangle,
\end{split}
\end{equation}
where the initial density operators was taken as
\begin{equation}\label{e4}
\rho(0)=\mid -1><-1\mid,~~ \text{i.e.,}~ \langle \langle -1;-1\mid\rho(0)\rangle \rangle =1.
\end{equation}
Using the determination of the dissipative operator $\vartheta^{0}_{C}$ (Eq.~\eqref{ae9})
and the results of the work~\cite{shirm2}, in which the calculating details the analogous expression for a one-nucleus spin system  are given,  
it is easy to obtain
\begin{equation}\label{e6098}
\begin{split}
\langle \langle -1;-1\mid \exp(-i\vartheta^{0}_{C}t)P^{0}\rho(0)\rangle \rangle = \exp(-2\gamma_{-1}t).
\end{split}
\end{equation}
Now let me examine the second and the third terms of the expression~\eqref{e39853}.  
For the second term  I have the result
\begin{equation}\label{e4671}
\begin{split}
&\langle \langle -1;-1\mid\exp(-i\vartheta^{0}_{C}t)D^{0}C^{0}P^{0}\rho(0)\rangle \rangle\\
&=\exp(-2\gamma_{-1}t) 
\frac{ \gamma_{N}^{2}H_{1}^{2}}{2}\Bigr{(}\int\limits_{0}^{\infty}\frac{f(\omega)}
{(\omega - \omega_{0} - i\varepsilon)^{2}}d\omega + c.c.\Bigl{)}.
\end{split}
\end{equation} 
The expression~\eqref{e4671} obtaining details are represented in Appendix A of this work.
Further, for the third term,  the following expression is correct~\cite{shirm2} 
\begin{equation}\label{e4657}
\begin{split}
&\sum\limits_{\nu\neq 0}\langle \langle -1; -1\mid C^{\nu}\exp(-i\vartheta^{\nu}_{C}t)P^{\nu}D^{\nu}P^{0}\rho(0)\rangle\rangle  \\
&=\frac{\gamma_{N}^{2}H_{1}^{2}}{2}\Big{(}\int\limits_{0}^{\infty}f(\omega)\frac{\exp(-i(\omega -\omega_{0})t)}
{(\omega - \omega_{0} - i\varepsilon)^{2}}d\omega + c.c.\Bigl{)}.
\end{split}
\end{equation} 
The integration in the expressions~\eqref{e4671},~\eqref{e4657} is carried out in the  investigated kinematic region, with the function $f(\omega)$ satisfing the condition
\begin{equation}\label{funct}
\int\limits_{0}^{\infty}f(\omega)d\omega = 1.
\end{equation}
Comparing the results~\eqref{e4671} and~\eqref{e4657} at the moment $t=0$, one can see 
\begin{equation}\label{e1946}
\begin{split}
&\langle \langle -1;-1\mid\exp(-i\vartheta^{0}_{C}t)D^{0}C^{0}P^{0}\rho(0)\rangle \rangle_{t=0} \\
&=\sum\limits_{\nu\neq 0}\langle \langle -1; -1\mid C^{\nu}\exp(-i\vartheta^{\nu}_{C}t)P^{\nu}D^{\nu}P^{0}\rho(0)\rangle \rangle_{t=0}
\end{split}
\end{equation}
whence the correct value for the matrix element at the initial moment of the time follows: $< -1\mid \rho (0)\mid -1> =1$.\\
Taking into account the relationships~\eqref{e1430},~\eqref{e39853} and~\eqref{e6098},~\eqref{e4671},~\eqref{e4657} I obtain the expression
\begin{equation}\label{e1093}
\begin{split}
&< -1\mid \rho (t)\mid -1> \\
&=\exp(-2\gamma_{-1}t) -\exp(-2\gamma_{-1}t) 
\frac{ \gamma_{N}^{2}H_{1}^{2}}{2}\Bigr{(}\int\limits_{0}^{\infty}\frac{f(\omega)}
{(\omega - \omega_{0} - i\varepsilon)^{2}}d\omega + c.c.\Bigl{)}\\
&+ \frac{ \gamma_{N}^{2}H_{1}^{2}}{2}\Bigr{(}\int\limits_{0}^{\infty}f(\omega)\frac{\exp(-i(\omega -\omega_{0})t)}
{(\omega - \omega_{0} - i\varepsilon)^{2}}d\omega + c.c.\Bigl{)}.
\end{split}
\end{equation}
The integrate calculation features the expression~\eqref{e1093} are examined in Appendix B. Calculating the integrals finally results in
\begin{equation}\label{e1112}
\begin{split}
&< -1\mid \rho (t)\mid -1> \\
&=\exp(-2\gamma_{-1}t) +\exp(-2\gamma_{-1}t) \frac{\gamma_{N}^{2}H_{1}^{2}}{2\pi\delta ^{2}\omega_{0}}\Bigr{(}2\delta +\pi \omega_{0} + 2\omega_{0}\arctan(\frac{\omega_{0}}{\delta})\Bigl{)}\\
&+\frac{\gamma_{N}^{2}H_{1}^{2}}{\delta\pi\omega_{0}^{2}t}(A(t)\sin(\omega_{0}t)+B(t)\cos(\omega_{0}t))-\exp(-\delta t)\frac{\gamma_{N}^{2}H_{1}^{2}}{\delta^{2}}.
\end{split}
\end{equation}
In the expression~\eqref{e1112} 
\begin{equation}\label{e7}
\begin{split}
A(t) = \int\limits_{0}^{\infty}d\xi\exp(-\xi)
\frac{a^{4}(a^{4}(1+b) - a^{2}(1+6b)\xi^{2}+b\xi^{4})}{D(\xi)},
\end{split}
\end{equation}
\begin{equation}\label{e8}
\begin{split}
B(t) =  \int\limits_{0}^{\infty}d\xi\exp(-\xi)
\frac{-a^{7}(2+4b)\xi + 4a^{5}b\xi^{3}}{D(\xi)}.
\end{split}
\end{equation}
\begin{equation}\label{e9}
\begin{split}
&D(\xi)=a^{8}(1+b)^{2} +a^{6}(2+b(2+4b))\xi^{2}+a^{4}(1+b(-2+6b))\xi^{4}\\
&+a^{2}b(-2+4b)\xi^{6}+b^{2}\xi^{8};
\end{split}
\end{equation}
$a=\omega_{0}t$ and $b=(\omega_{0}/\delta)^{2}$.

\section{Results and discussion}
In Fig. 1, 2 the transition from the energy level $E_{3}$ to the level $E_{2}$ of the pair of the connected hydrogen nuclei $H(1)$ (mass number $A=1$) is investigated. The transition is accompanied by an $\omega_{0}=\Delta E_{23}$ energy quantum emitting. The evolution of the density matrix is examined for a long time scale (Fig. 1) and for a short time scale (Fig. 2). It is shown that the behavior of the complete density matrix~\eqref{e1112} (solid line) is slower than the exponential one~\eqref{e6098} (dashed line). In the asymptotic time region, $t\rightarrow +\infty$, it approaches to the latter. The complete density matrix is determined on the short time scale, where the deviation from the exponential, Markov evolution is seen (Fig 2). The behavior of the nuclear spin system is found depending on the initial dynamic parameters selection. For example, Fig. 1 presents the  calculations for the field $H_{1}=37$~Oe. In this case, the deviation from the exponential evolution is not considerable.
Fig. 3,4~present the calculation of the time evolution of the connected carbon nuclei C(13) (mass number $A=13$). The calculation is carried out for the  field $H_{1}=100$~Oe and for the field $H_{1}=150$~Oe  (Fig.3). In this case, the  behavior dynamics of the nuclear spin system coincides with the foregoing case. The interaction with the field $H_{1}=100$~Oe leads to a noticeable deviation  from the Markov evolution. In the case $H_{1}=150$~Oe the deviation is not considerable. 
On the short time scale, (Fig.4) the nuclear spin system behavior is slower than in the H(1) case.
In both cases, the calculations  were made  for: $\delta =10^{6}~ sec^{-1}$, $\theta=30^\circ$, $r=2\times 10^{-9}~cm$. The magnetic moments of the nuclei have the values: 2.7927 for H(1) and 0.702381 for C(13) (the values of the magnetic moments  are given in the units of the nuclear magneton). \\
The suppression of transitions between quantum states obtained on a short time scale in the course of the present work may be thought of as the so-called Zeno effect occuring as a measurement result in quantum physics. At present, Zeno effect is discussed widely in the literature (see, for example, the works~\cite{Zen1}-~\cite{Xiao}). Sudarshan and Misra's pioneer work ~\cite{Zen1} showed that continuous observation of the decay process makes the latter impossible as  
the observation influences the decay. In other words,  the decay slows down under the observation and in the case of continuous, ideal measurement it becomes impossible. Thus, Zeno effect contains a subjective, anthropomorphic element as quantum theory describes the fact that we can speak about a quantum world after taking measurement rather than quantum world itself. The result depends on the observer and on observation conditions. 
This paradoxical result was subsequently examined in connection with the problem of quantum transitions in atoms~\cite{Zen8}. The theoretical aspects of the paradox are discussed in the monograph~\cite{Zeno} (see also  Khalfin~\cite{Zen3} and the work~\cite{Zen9}). \\
Quantum Zeno effect was investigated by the Brussels-Austin group in the works~\cite{Zen4}~-~\cite{Zen6}. In the work~\cite{Zen4}, the system of harmonic oscillators coupled with an external field was examined. Zeno effect was discussed in connection with the problem of quantum decoherence. It has been shown that quantum Zeno time serves as a lower bound for the decoherent process. In the work~\cite{Zen5}, a particle coupled with a  field in the thermodynamic limit on the basis of the complex spectral representation of the Liouvillian was discussed. The results show the time derivative of the photon number density asymptotically vanished at $t\rightarrow 0$. It was shown that the influence of Zeno effect is noticeable and determines the subsequent system dynamics. The short-time behaviour of the survival probability in the frame of the Friedrichs model was investigated in the work~\cite{Zen6}. The problem was associated with a photodetachment process, quantum dot, and a hydrogen atom. It was discovered that Zeno effect occurs in a very short time region. \\
In connection with the quantum computation problems, the studies dealing with creating quantum systems for which Zeno time region is long and which make quantum calculation possible are of great importance. In this connection, the work of Center for Quantum Computation, University of Oxford is noteworthy. The work~\cite{Xiao} presented the  preliminary experimental results demonstrating the quantum Zeno effect in NMR for H(1), C(13) nuclear spin systems. The specific features of the future NMR experiments, in which Zeno effect can be demonstrated, are discussed. 
The authors  state,  that "Zeno effect in NMR experiments has not previously been explored, probably as a result of the dificulty of performing true quantim measurement."

\section{Conclusion}
The investigation of the evolution of the two-nucleus spin systems interacting with the magnetic field in the frame of the complex spectral representation of the Liouvillian gives the rigorous description of the irreversibility.  The short time (Zeno) region determines the memory effect on the quantum level. In the approach, Zeno effect is considered as an objective dynamic process taking place in a quantum systems on a short time scale  without using anthropomorphic elements. The study of the two-nucleus spin systems reveals the significant role of the effect which, however, depends on the characteristics of the magnetic field interacting with the nuclei. The investigation of the relaxation process for the two-nucleus spin systems demonstrates 
the utilization of the Brussels-Austin formalism in a practical problem that may be tested in experiment.

{\bf Acknowledgements}\\
I am grateful to  Dr. D. V.~Shulga  for the helpful
discussion and Dr. A. A. Goy, Dr. A. V.~Molochkov~for the
support of this work.

\setcounter{equation}{0}
\def\theequation{A.\arabic{equation}}
{\bf Appendix A. The second approximation }\\
Let me examine the second term of the expression~\eqref{e39853}\\ $\langle \langle -1;-1\mid\exp(-i\vartheta^{0}_{C}t)D^{0}C^{0}P^{0}\rho(0)\rangle \rangle$. The operators 
$C^{0}$, $D^{0}$ have the form
\begin{equation}\label{e32531}
\begin{split}
C^{0}=\sum\limits_{\mu\neq 0}P^{\mu}\frac{-1}{w^{\mu}-i\varepsilon}L_{I}P^{0},
\end{split}
\end{equation}
\begin{equation}\label{d221}
D^{0}=\sum\limits_{\mu\neq 0}P^{0}L_{I}\frac{1}{-w^{\mu}+i\varepsilon}P^{\mu}.
\end{equation}
In our model, the projection operators $P^{0}$, $P^{\mu}$ ($\mu\neq 0$) are determined as follows 
\begin{equation}\label{e156}
\begin{split}
P^{0}=\mid -1;-1\rangle \rangle\langle \langle -1;-1\mid +
\frac{1}{\Omega}\int\limits_{-\infty}^{\infty}\mid 0 ~\omega ; 0~ \omega\rangle \rangle\langle \langle 0 ~\omega ; 0 ~\omega\mid d\omega,
\end{split}
\end{equation}
\begin{equation}\label{e256}
\begin{split}
P^{\mu}=\int\limits_{-\infty}^{\infty}\mid -1;0\omega\rangle \rangle\langle \langle -1;0\omega\mid d\omega + 
\int\limits_{-\infty}^{\infty}\mid 0\omega;-1\rangle \rangle\langle \langle 0\omega;-1\mid d\omega,
\end{split}
\end{equation}
where the value $\Omega$ is designated  so that $\frac{\delta(0)}{\Omega}=1$ and $\delta(0)$ is the delta-function.
The expressin~\eqref{e4671} is obtained in the limit $\Omega\rightarrow\infty$. In this case 
\begin{equation}\label{e3562}
\begin{split}
&\langle \langle -1;-1\mid\exp(-i\vartheta^{0}_{C}t)D^{0}C^{0}P^{0}\rho(0)\rangle \rangle \\
&=\sum\limits_{\mu\neq 0}\langle \langle -1;-1\mid\exp(-i\vartheta^{0}_{C}t)\mid -1;-1\rangle \rangle\langle \langle -1;-1\mid L_{I}\frac{1}{-w^{\mu}+i\varepsilon}P^{\mu}\\
&\times\frac{-1}{w^{\mu}-i\varepsilon}L_{I}\mid -1;-1\rangle \rangle\langle \langle -1;-1\mid\rho(0)\rangle \rangle. 
\end{split}
\end{equation}
For the first cofactor of~\eqref{e3562}, I have the result~\eqref{e6098}. For the second cofactor, taking into account the determinations 
\begin{equation}\label{e234}
L_{I}=H_{I}\times 1 -1\times H_{I},~~(A\times B)C=ACB,
\end{equation}
\begin{equation}\label{e345}
\begin{split}
\langle \langle\alpha ;\beta \mid L_{I}\mid\alpha ' ;\beta '\rangle \rangle = <\alpha\mid H_{I}\mid\alpha '>\delta_{\beta '\beta } - \delta_{\alpha\alpha '}<\beta '\mid H_{I}\mid\beta>,
\end{split}
\end{equation}
where $\delta_{\beta '\beta }$ ($\delta_{\alpha\alpha '}$) is delta-function if the indices $\beta '$, $\beta$ ($\alpha$, $\alpha '$)
correspond to the continuous spectrum and it is Kronecker's symbol if the indices correspond to the discrete spectrum,
it is possible to write down 
\begin{equation}\label{e35623}
\begin{split}
&\sum\limits_{\mu\neq 0}\langle \langle -1;-1\mid L_{I}\frac{1}{-w^{\mu}+i\varepsilon}P^{\mu}
\frac{-1}{w^{\mu}-i\varepsilon}L_{I}\mid -1;-1\rangle \rangle \\
&=\sum\limits_{\rho}\Bigr{(}\frac{<-1\mid H_{I}\mid\rho><\rho\mid H_{I}\mid -1>}{(E_{\rho}-E_{3}-i\varepsilon)^{2}} \\
&+\frac{<\rho\mid H_{I}\mid -1><-1\mid H_{I}\mid \rho>}{(E_{3}-E_{\rho}-i\varepsilon)^{2}}
\Bigl{)}\\
&=\frac{\gamma_{N}^{2}H_{1}^{2}}{2}\Bigr{(}\int\limits_{0}^{\infty}\frac{f(\omega)}
{(\omega - \omega_{0} - i\varepsilon)^{2}}d\omega + c.c.\Bigl{)}.
\end{split}
\end{equation} 
Here,  index $\rho $ corresponds to the state $\mid 0, \omega>$,  $E_{\rho} =E_{2}+\omega$, $\omega_{0}=E_{3}-E_{2}$.
In~\eqref{e35623}, the sum determines the summation (integration) over all discrete (continuous) indices. The results~\eqref{e6098} and~\eqref{e35623} lead to the expression~\eqref{e4671}.

\setcounter{equation}{0}
\def\theequation{B.\arabic{equation}}
{\bf Appendix B. Calculation of the integrals }\\
I examine the second integral of the expression~\eqref{e1093}
\begin{equation}\label{e122322}
\begin{split}
\int\limits_{0}^{\infty}f(\omega)\frac{\exp(-i(\omega -\omega_{0})t)}
{(\omega - \omega_{0} - i\varepsilon)^{2}}d\omega. 
\end{split} 
\end{equation}
Since, $t>0$, we have to deform the contour  of integration $\Gamma_{R}$ in the lower half plane~\cite{dec}. The application of the residues theorem leads to
\begin{equation}\label{e1223}
\begin{split}
&\int\limits_{\Gamma_{R}}\exp(-i(z-\omega_{0})t)F(z)dz=-2\pi i\sum\limits_{z=z_{i}}Res~[\exp(-i(z-\omega_{0})t)F(z)],
\end{split} 
\end{equation}
where 
\begin{equation}\label{e1234}
F(z) = \frac{f(z)}
{(z - \omega_{0} )^{2}}
\end{equation}
and the sum comes from the poles $z_{i}$ of the function $F(z)$.  
In the expression~\eqref{e1223}
\begin{equation}\label{e1224}
\begin{split}
&\int\limits_{\Gamma_{R}}\exp(-i(z-\omega_{0})t)F(z)dz= \int\limits_{0}^{R}\exp(-i(\omega-\omega_{0}) t)F(\omega)d\omega \\
&+\int\limits_{\gamma_{R}}\exp(-i(z-\omega_{0})t)F(z)dz + \int\limits_{-iR}^{0}\exp(-i(z-\omega_{0})t)F(z)dz.
\end{split}
\end{equation}
If $f(z)$ is the Lorentzian distribution~\eqref{e867} then,
\begin{equation}\label{e765}
\lim\limits_{R\rightarrow \infty}\int\limits_{\gamma_{R}}\exp(-izt)F(z)dz \rightarrow 0 .
\end{equation}
In the lower half plane the function $F(z)$ has one pole $z=\omega_{0}-i\delta$. In this case 
\begin{equation}\label{e1229}
\begin{split}
-2\pi i\sum\limits_{z=z_{i}}Res[\exp(-i(z-\omega_{0})t)F(z)]=-\frac{\exp(-\delta t)}{\delta^{2}}.
\end{split} 
\end{equation}
Using the  result obtained in the limit $R\rightarrow \infty$ we can write down 
\begin{equation}\label{908}
\int\limits_{0}^{\infty}\exp(-i(\omega-\omega_{0}) t)F(\omega)d\omega =  \int\limits_{0}^{-i\infty}\exp(-i(z-\omega_{0})t)F(z)dz-\frac{\exp(-\delta t)}{\delta^{2}}.
\end{equation}
The replacement  $z=-iy$,  makes it possible to reduce the result~\eqref{908} to the form
\begin{equation}\label{9081}
\begin{split}
&\int\limits_{0}^{\infty}\exp(-i(\omega-\omega_{0}) t)F(\omega)d\omega = -i \int\limits_{0}^{\infty}\exp(-i(-iy-\omega_{0})t)F(-iy)dy \\
&-\frac{\exp(-\delta t)}{\delta^{2}}.
\end{split}
\end{equation}
Then the replacement $y=\xi /t$, where $\xi \geq 0$ leads to the final expression
\begin{equation}\label{e10075}
\begin{split}
&\int\limits_{0}^{\infty}f(\omega)\frac{\exp(-i(\omega -\omega_{0})t)}
{(\omega - \omega_{0} - i\varepsilon)^{2}}d\omega + c.c.\\
&=\frac{2}{\delta\pi\omega_{0}^{2}t}(A(t)\sin(\omega_{0}t)+B(t)\cos(\omega_{0}t))-2\frac{\exp(-\delta t)}{\delta^{2}}.
\end{split}
\end{equation}
It is necessary to note, that the result  for the integral~\eqref{e122322} in the foregoing work~\cite{shirm2} was obtained without taking into account the residue~\eqref{e1229}. This was made because the role of the latter in the investigated kinematic region was not essential.\\    
The integral
\begin{equation}\label{e12232}
\begin{split}
\int\limits_{0}^{\infty}\frac{f(\omega)}
{(\omega - \omega_{0} - i\varepsilon)^{2}}d\omega
\end{split} 
\end{equation}
can be obtained from the expression~\eqref{908} 
for the case $t=0$ 
\begin{equation}\label{9082}
\int\limits_{0}^{\infty}F(\omega)d\omega =  \int\limits_{0}^{-i\infty}F(z)dz-\frac{1}{\delta^{2}}.
\end{equation}
Using the "Maxima" program and taking into account the complex conjugion, I then have the result
\begin{equation}\label{e4653}
\begin{split}
\int\limits_{0}^{\infty}\frac{f(\omega)}
{(\omega - \omega_{0} - i\varepsilon)^{2}}d\omega + c.c.=-\frac{2\delta +\pi \omega_{0} +2 \omega_{0}\arctan(\frac{\omega_{0}}{\delta})}{\pi\delta ^{2}\omega_{0}}.
\end{split}
\end{equation}

\newpage

\begin{figure}[h]
\includegraphics[height=76.5mm~~~~~~~~]{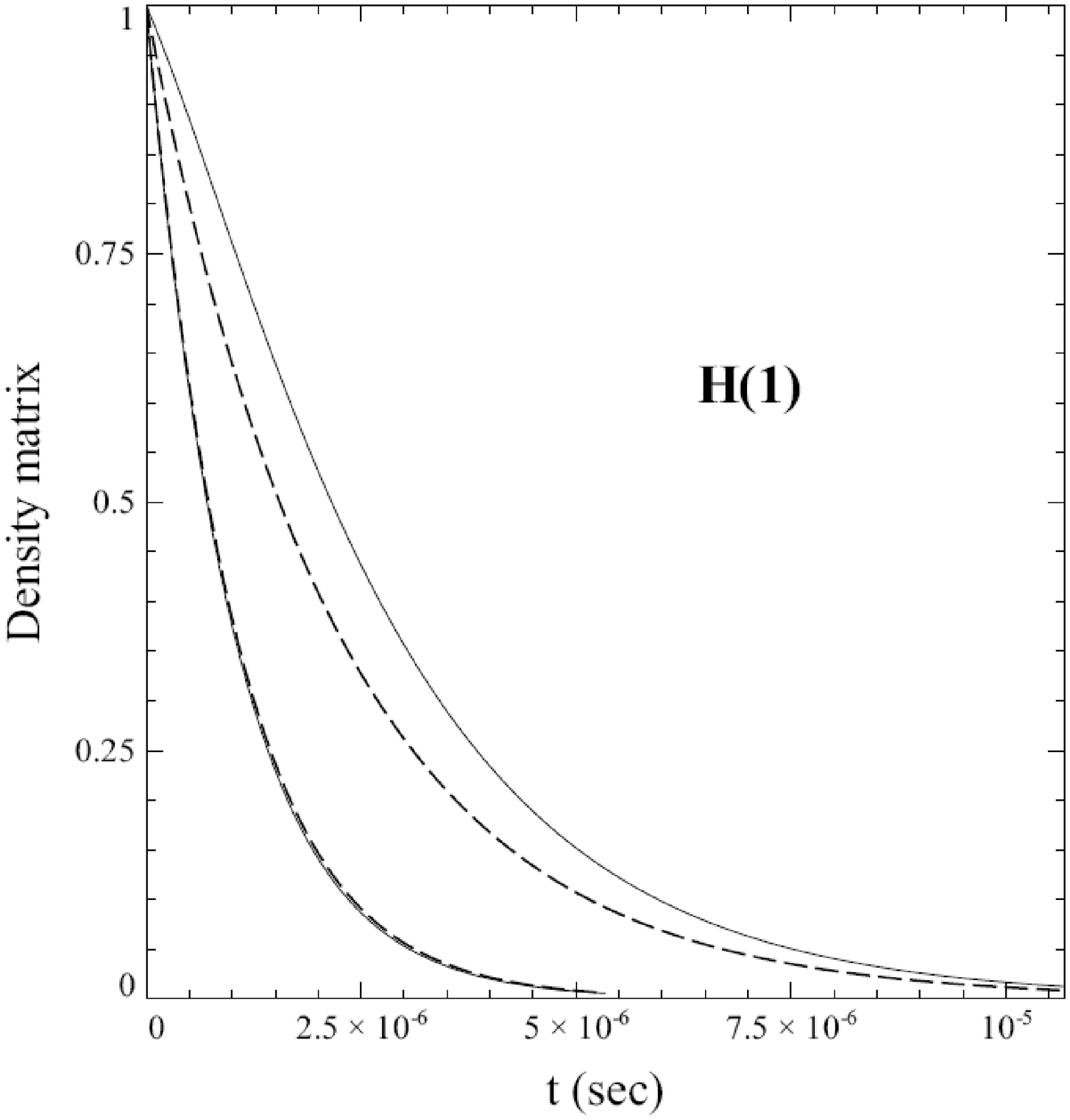}
\caption{}
\label{fig1}
\end{figure}
\hspace{5mm}

\begin{figure}[h]
\includegraphics[height=76.5mm]{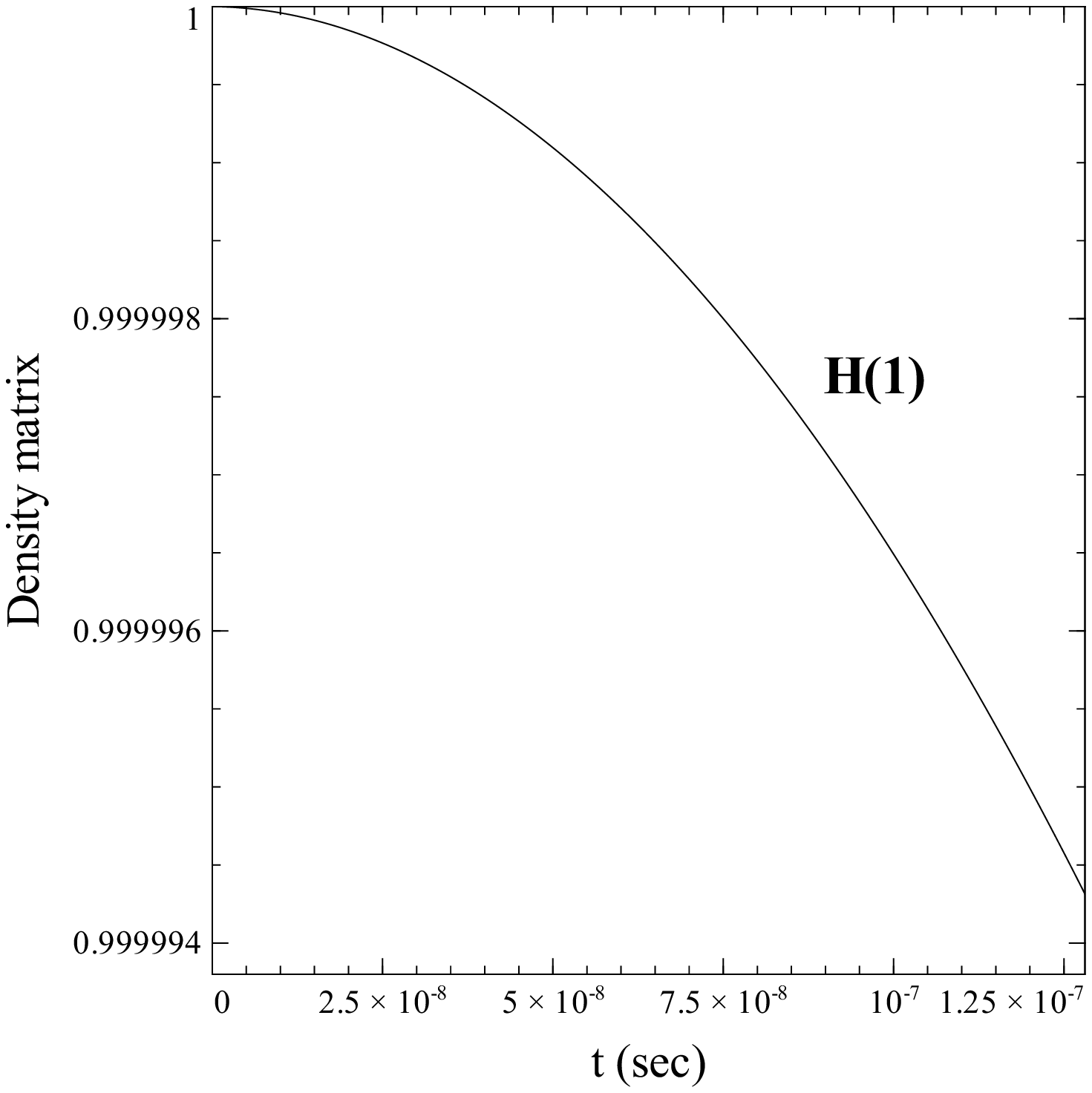}
\hspace{5mm}
\includegraphics[height=76.5mm]{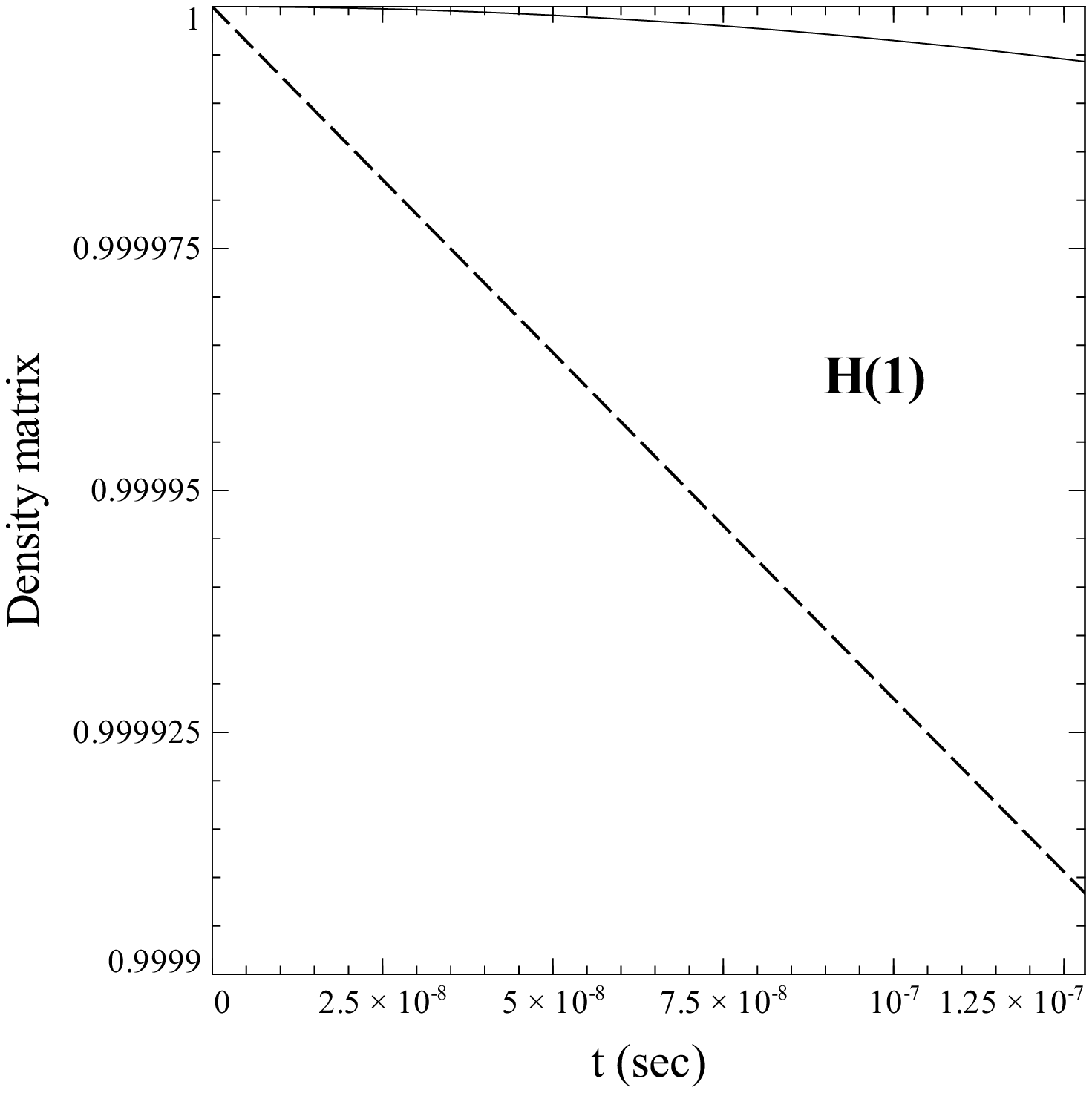}
\caption{}
\label{fig2}
\end{figure}

\newpage
\begin{figure}[h]
\includegraphics[height=76.5mm~~~~~~~~]{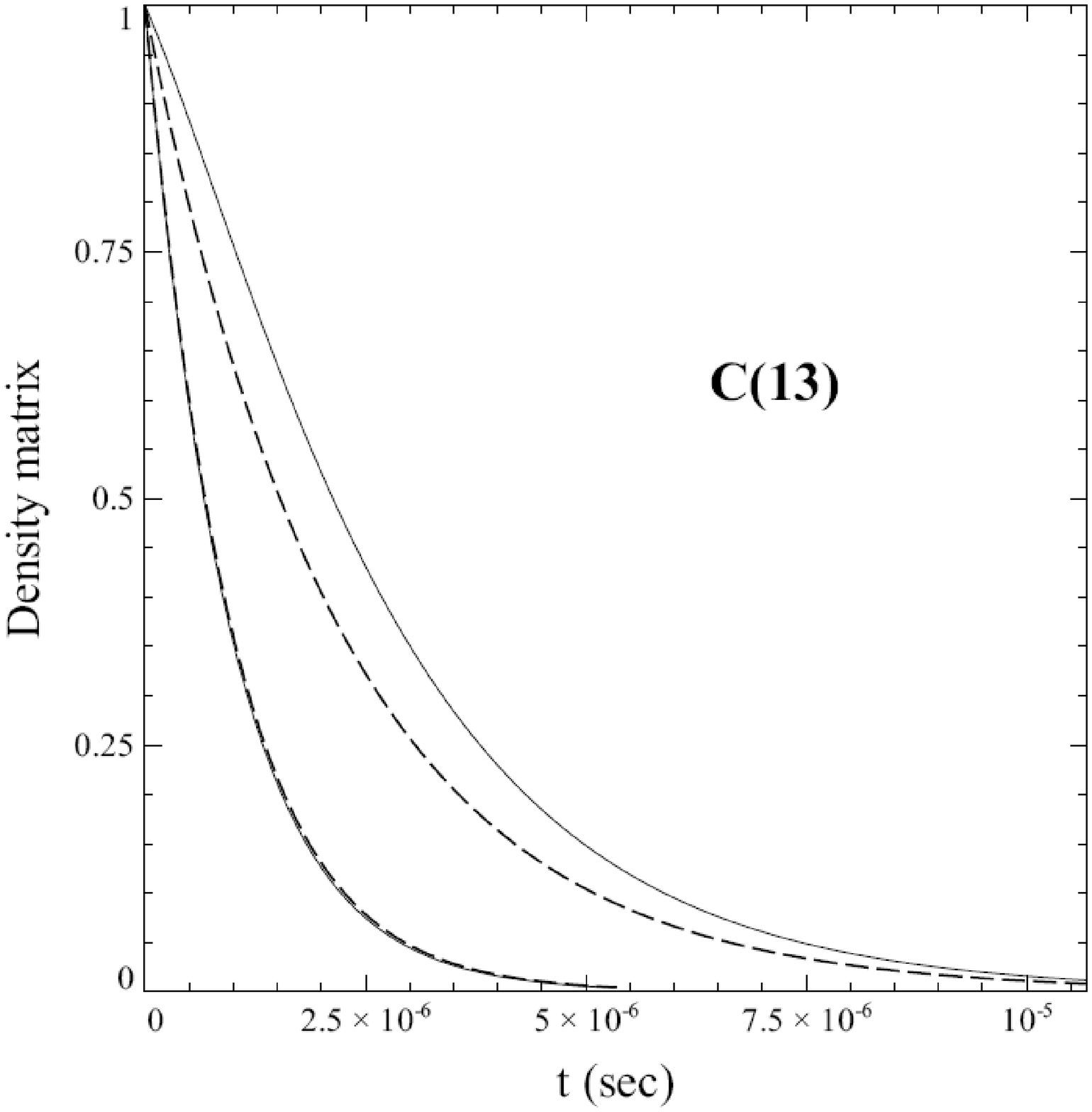}
\caption{}
\label{fig3}
\end{figure}

\begin{figure}[h]
\includegraphics[height=76.5mm]{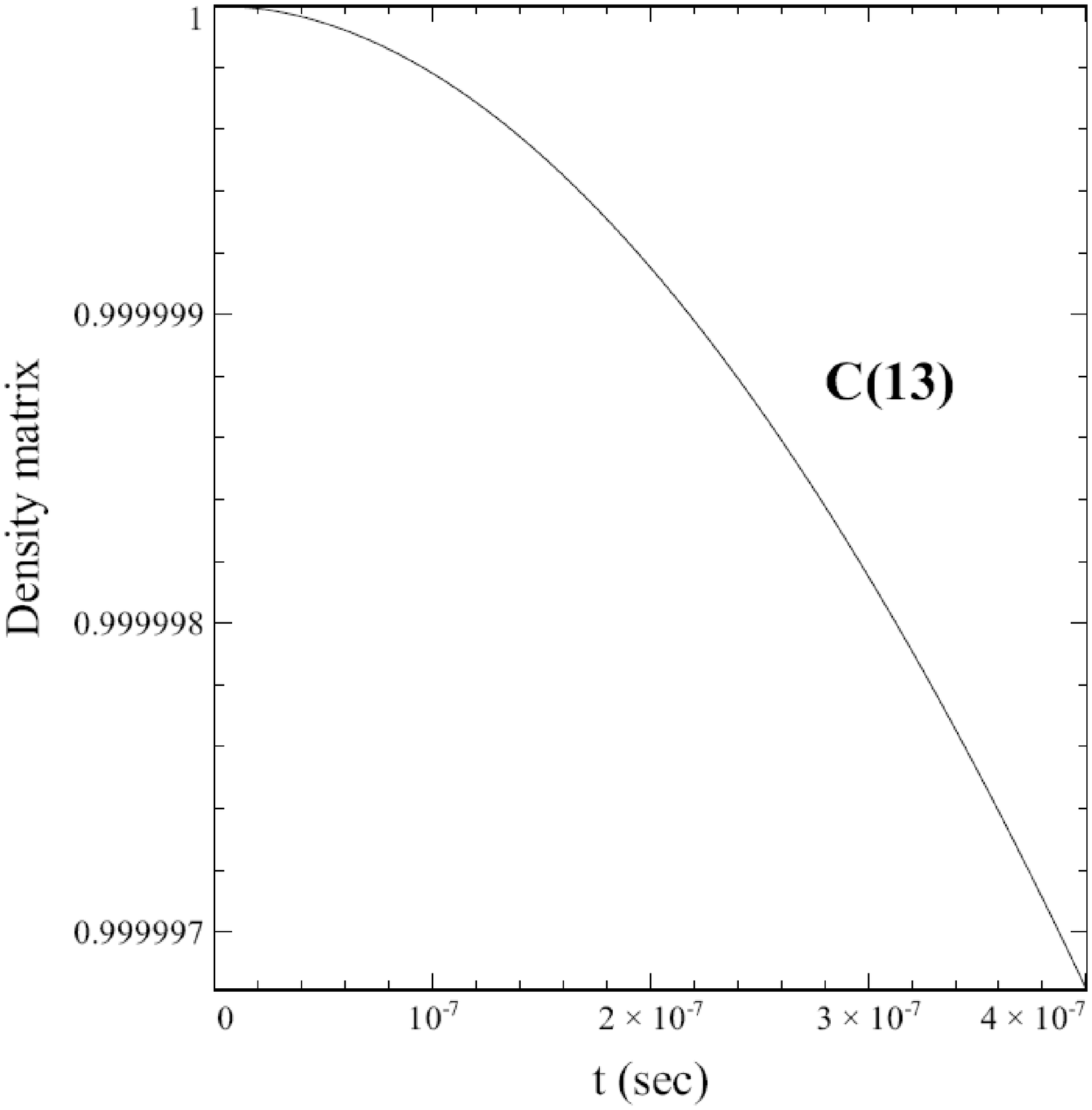}
\hspace{5mm}
\includegraphics[height=76.5mm]{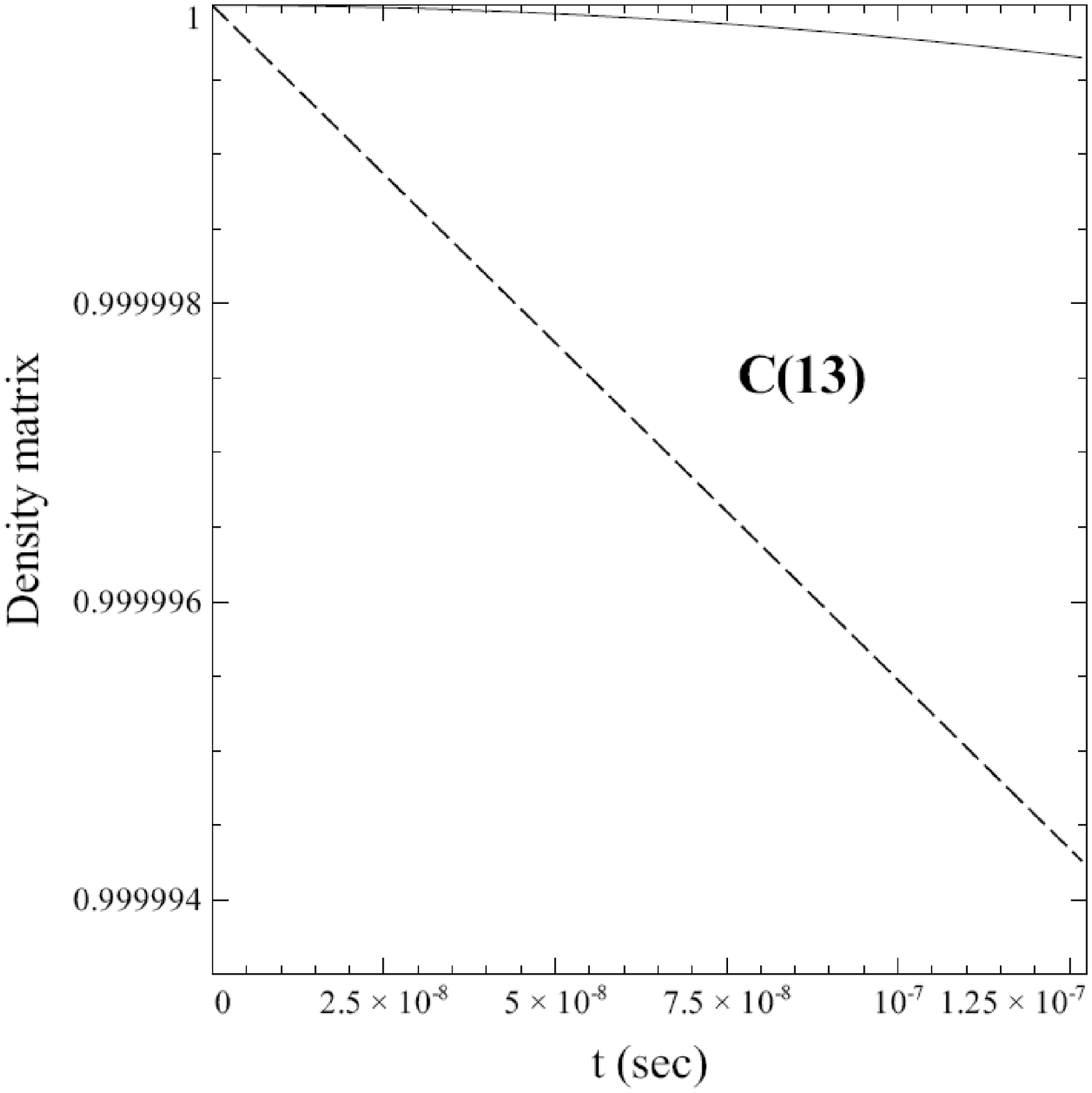}
\caption{}
\label{fig4}
\end{figure}

{\bf Figure Legends}

\vspace*{3mm} Fig.1:  H(1), $H_{z}=10^{4}~Oe$, $H_{1}=25~Oe$ and $H_{1}=37~Oe$; solid line is for complete density matrices~\eqref{e1112}, dashed line is exponential one~\eqref{e6098}.

\vspace*{4mm} Fig.2:  H(1), $H_{z}=10^{4}~Oe$, $H_{1}=1~Oe$; the designation of the lines as in Fig. 1.

\vspace*{4mm} Fig.3:  C(13), $H_{z}=10^{4}~Oe$, $H_{1}=100~Oe$ and $H_{1}=150~Oe$; the designation of the lines as in Fig. 1.

\vspace*{4mm} Fig.4:  C(13), $H_{z}=10^{4}~Oe$, $H_{1}=1~Oe$; the designation of the lines as in Fig. 1.

\end{document}